\documentclass[onecolumn,iop,apj]{emulateapj}   
\usepackage{graphicx}        % For eps figures, newer & more powerful
\usepackage{amssymb}         % useful mathematical symbols
\usepackage[usenames,dvipsnames]{color}          % For color text: \color command
\definecolor{darkblue}{rgb}{0,0,0.5}

\DeclareFontFamily{OT1}{pzc}{}
\DeclareFontShape{OT1}{pzc}{m}{it}%
             {<-> s * [1.1500] pzcmi7t}{}
\DeclareMathAlphabet{\mathscr}{OT1}{pzc}%
                                 {m}{it}

\usepackage{amsmath}
\newcommand{\half}{{\textstyle\frac{1}{2}}}

\newcommand{\rd}{\mathrm{d}}

\newcommand{\pderiv}[2]{\frac{\partial#1}{\partial#2}}

\newcommand{\deriv}[2]{\frac{\rd#1}{\rd#2}}
\newcommand{\derivd}[2]{\frac{\rd^2#1}{\rd#2^2}}

\newcommand{\x}{\mathbf{x}}

\newcommand{\vdot}{{\boldsymbol{\cdot}}}

\newcommand{\grad}{\mbox{\boldmath$\nabla$}}
\newcommand{\bxi}{\mbox{\boldmath$\xi$}}

\newcommand{\thth}{\hspace{1.5pt}}

\newcommand\Div{\grad\vdot\thth}

\renewcommand{\bv}{Brunt-V\"ais\"al\"a}

\newcommand{\ri}{{i}}

\renewcommand{\leq}{\leqslant}  \renewcommand{\le}{\leqslant}
  \renewcommand{\ge}{\geqslant}

\allowdisplaybreaks[1]

\begin{document}

%%%%%%%%%%%%%%%%%%%%%%%%%%%%%%%%%%%%%%%%%%%%%%%%%%%%%%%%%%%%%

%\title{How Compact Acoustic Sources in the Solar Convection Zone Hide Themselves}
\title{Near- and Far-Field Response to Compact Acoustic Sources in Stratified Convection Zones}

\author{Paul S.~Cally}

\affil{Monash Centre for Astrophysics and School of Mathematical Sciences,\\ Monash University, Clayton, Victoria 3800, Australia}
  \email{paul.cally@monash.edu}

\shortauthors{P.S. Cally}

%\shorttitle{Hiding Compact Acoustic Sources}
\shorttitle{Near- and Far-Field Response to Compact Acoustic Sources}

\begin{abstract}
The role of the acoustic continuum associated with compact sources in the Sun's interior wave field is explored for a simple polytropic model. The continuum produces a near-field acoustic structure -- the so-called `acoustic jacket' -- that cannot be represented by a superposition of discrete normal modes. Particular attention is paid to monochromatic point sources of various frequency and depth, and to the surface velocity power that results, both in the discrete $f$- and $p$-mode spectrum and in the continuum. It is shown that a major effect of the continuum is to heal the surface wave field produced by compact sources, and therefore to hide them from view.
It is found that the continuous spectrum is not a significant contributor to observable inter-ridge seismic power.
\end{abstract}

\keywords{Sun: helioseismology -- Sun: oscillations}

%%%%%%%%%%%%%%%%%%%%%%%%%%%%%%%%%%%%%%%%%%%%%%%%%%%%%

\section{Introduction}
Acoustic oscillations in the solar convection zone are believed to be turbulently excited in a shallow sub-surface layer \citep{GolMurKum94aa,SteNor01aa,SamNorSte03aa}. In acoustic glories surrounding active regions, local helioseismic imaging reveals very compact bead-like emission in the 5-6 mHz band \citep{DonLinBra00aa}, suggesting that at least a component of wave excitation is coming from effectively point-like sources with spatial extent far smaller than the typical wavelengths they generate. It is unclear whether these sources are associated with thin magnetic flux tubes, but such small scale structure does suggest the possibility.

The thin magnetic flux tubes that ubiquitously thread the solar surface are known to be potential acoustic sources \citep{Cal86aa}. Essentially transverse slow magneto-acoustic waves travelling down these tubes into the Sun's interior exhibit ever smaller wavelengths as the Alfv\'en speed decreases with depth \citep{calbogzwe94aa,BogHinCal96aa}. Matching transverse velocity across the tube boundary therefore results in the generation of acoustic waves in the surrounding non-magnetic atmosphere with very large vertical wave numbers $k_z$. The horizontal wavenumber $k_h$, given by the acoustic dispersion relation $k_h^2=(\omega^2-\omega_c^2)/c^2-k_z^2$ (where $\omega_c$ is the acoustic cutoff frequency), is therefore necessarily imaginary for given frequency $\omega$ and sound speed $c$ such that $k_z^2>(\omega^2-\omega_c^2)/c^2$. This manifests as the `acoustic jacket' \citep{bogcal95aa}, a horizontally evanescent sheath surrounding the flux tube. The acoustic jacket is distinct from the usual discrete spectrum of $f$- and $p$- modes familiar in helioseismology, and mathematically is associated with an additional continuous spectrum.

%The dispersion relation argument makes it clear that a necessary prerequisite for the acoustic jacket is a specified and sufficiently large vertical wavenumber. This is a natural feature of flux tubes where structure is imposed along their length by internal MHD (magnetohydrodynamic) wave modes. Isolated (non-magnetic) sources however do not have to abide by any such vertical structure. A point source could act as a monopole exciting purely radial oscillations (sufficiently close to the source) satisfying $\omega^2-\omega_c^2=c^2 |\k|^2$, in which case no jacket need arise. Multipole sources however may behave differently, subject to orientation.

The acoustic jacket is widely known but difficult to work with, though of course it appears naturally in grid-based simulations. Its mathematics is rather challenging, and not easily implemented in  modally-based calculations. Indeed, it has become common practice to artificially truncate the atmosphere at some depth so as to introduce a new unphysical discrete set of modes that can partially mimic the continuum of jacket modes \citep{BarCal00aa}, or even to ignore the acoustic jacket totally (with apologies) deeming it too difficult \citep[e.g.,][]{JaiHinBra09aa}.

Compact sources are easily introduced into simulation codes, and acoustic jackets will appear naturally, though spatial resolution is an issue if they are thin. However, there is some value in understanding them from an analytic point of view to better alert us to their potential consequences. In that vein, the purpose of this paper is to revisit the basic jacket mode theory and show how vertical line sources may be constructed with it. This is then used to construct full 3D cylindrical or 2D cartesian wave fields originating from (specifically) point sources, which naturally exhibit a rich spectrum of high-$k_z$ oscillations. We pay particular attention to the observable $f$- and $p$-mode surface power as functions of source depth and frequency.
%, and investigate the extent to which buried compact sources can be `seen' or at least resolved at the surface by imaging techniques based on optical analogies \citep[\emph{e.g.} Acoustic Holography,][]{LinBra00aa}. 
It is found that the continuum acts to `heal' the surface signature of point sources by `filling in' their image in the discrete-spectrum surface velocity field.

%%%%%
\section{Eigenfunction Expansion of Linear Waves in a Truncated Polytrope}
\subsection{Discrete and Continuous Spectra}
Details of the development of the wave equations in a truncated adiabatic polytropic atmosphere may be found in \citet[Section 2]{bogcal95aa}. We summarise them here.

Consider an adiabatic plane polytrope of index $m=1/(\gamma-1)$, where $\gamma$ is the ratio of specific heats, such that the equilibrium density and pressure vary with depth $z$ according to
\begin{equation}
\rho(z)=\rho_0\,\left(\frac{z}{z_0}\right)^m, \qquad p(z)=p_0\,\left(\frac{z}{z_0}\right)^{m+1},
\end{equation}
where $p_0=\rho_0 g z_0/(m+1)$, $g$ is the gravitational acceleration, and $z_0$ is the truncation depth, \emph{i.e.} $z_0\leq z<\infty$. A stress-free boundary condition $\Div\bxi=0$ is imposed at $z_0$, where $\bxi$ is the fluid displacement vector, and regularity is enforced at $z=\infty$.

This atmosphere supports both $f$- and $p$-modes, but not $g$-modes as it is adiabatic (the {\bv} frequency $N=0$). The vorticity therefore vanishes, and hence the displacement is potential, $\bxi=z_0^2\grad\Phi$ (the factor of $z_0^2$ has been included to make $\Phi$ dimensionless). Adopting the convenient non-dimensionalization $s=z/z_0\ge1$, the wave equation takes the form
\begin{equation}
\left(\tilde\nabla^2+\frac{m}{s}\pderiv{}{s}+\frac{\nu^2}{s}\right)\Phi(\x_\perp,s) = 0
\label{waveeqn3D}
\end{equation}
where $\tilde\grad=z_0\grad$ and $\nu=\omega\sqrt{m z_0/g}$ is dimensionless frequency. The velocity potential $\Phi$ may also be understood as a simple multiple of the Eulerian pressure perturbation,
\begin{equation}
p'=\gamma p_0 \nu^2 s^m \Phi  \,. \label{p'}
\end{equation}
The boundary conditions are
\begin{gather}
\left(\pderiv{}{s}+\frac{\nu^2}{m}\right)\Phi=0, \quad \mbox{as $s\to1^+$,}  \label{BC1}\\[6pt]
\Phi\to0, \quad \mbox{as $s\to+\infty$.}  \label{BC2}
\end{gather}

Equation (\ref{waveeqn3D}) with boundary conditions (\ref{BC1}-\ref{BC2}) may be analysed using separation of variables. The vertical behaviour is found to satisfy the ordinary differential eigenvalue equation
\begin{equation}
\left(\derivd{}{s}+\frac{m}{s}\deriv{}{s}+\frac{\nu^2}{s}+\lambda\right)\Phi_\parallel = 0
\label{waveeqn}
\end{equation}
where a monochromatic potential
\begin{equation}
\Phi(x,s,t)=\Phi_\parallel(s)\exp[\ri(k x-\omega t)]
\end{equation}
is assumed, with $x$ being horizontal distance and $t$ time and $\lambda=-k^2 z_0^2$. For simplicity we have assumed a two-dimensional ($x$-$z$) scenario; cylindrical 3D is easily addressed using Bessel functions in horizontal distance rather than a complex exponential. More general time dependence may be recovered by Fourier superposition.

Equation (\ref{waveeqn}) with the stated boundary conditions may be solved in terms of Whittaker functions \citep[Ch. 13]{DLMF}. Specifically, we define
\begin{gather}
\phi(s;\lambda) = s^{-1/2-\mu}
\left[2\mu M_{\kappa,-\mu-1}(\zeta) M_{\kappa,\mu}(s\zeta) +
\frac{(1/2+\mu)^2-\kappa^2}{2(1+\mu)(1+2\mu)^2}M_{\kappa,\mu+1}(\zeta)M_{\kappa,-\mu}(s\zeta)\right] , \label{phi}\\[6pt]
\psi(s;\lambda) = s^{-1/2-\mu} W_{\kappa,\mu}(s\zeta), \label{psi}
\end{gather}
where $\mu=(m-1)/2$, $\kappa=\ri\,\nu^2/2\sqrt{\lambda}$, and $\zeta=-2\ri\sqrt{\lambda}$. In numerical examples, we shall use $\mu=1/4$ throughout, corresponding to $m=3/2$, \emph{i.e.} $\gamma=5/3$. It is assumed that $\kappa$ and $\zeta$ are defined on $0\le\arg\lambda<2\pi$, so that both $\arg\kappa$ and $\arg\zeta$ lie in $[-\pi/2,\pi/2)$. (Computationally, it is convenient to express these as $\zeta=2/\sqrt{-1/\lambda}$ and $\kappa=\nu^2/\zeta$ where the square root applies the usual cut along the negative real axis.) These two solutions of Equation (\ref{waveeqn}) satisfy the boundary conditions at $s=1$ and $\infty$ respectively. Their Wronskian
\begin{equation}
\mathcal{W}_\lambda[\phi,\psi](s) = 2\ri\mu s^{-1-2\mu} \left(\sqrt{\lambda}-\sqrt{\lambda_0}\right) W_{\kappa,\mu+1}(\zeta)
\end{equation}
possesses simple zeros at $\sqrt{\lambda_0}\equiv \ri\,\nu^2/(1+2\mu)$, corresponding to the $f$-mode $\omega^2=kg$, and an infinite sequence $\lambda_n$ ($n=1,2,3,\ldots$) of $p$-modes such that $\lambda_0<\lambda_1<\lambda_2<\cdots<0$. These become simple poles in the Green's function, and form the familiar discrete spectrum.

\citet{bogcal95aa} go on to show that the $f$- and $p$-modes are not complete. One must also include a continuous spectrum, representing the jacket modes. Specifically, a suitably integrable arbitrary function $f(s)$ may be expanded thus:
\begin{equation}
f(s) = \sum_{n=0}^\infty F_n\, \psi_n(s) + \int_0^\infty \mathcal{F}_\lambda\, \phi(s;\lambda)\,d\lambda   \label{expand}
\end{equation}
where $\psi_n(s) =\psi(s;\lambda_n)$ and
\begin{align}
F_n &= A_n\, \int_1^\infty r^{1+2\mu} \psi_n(r)f(r)\,dr, \quad\mbox{($n=0,1,2,\ldots$)},\label{Fn}\\[6pt]
\mathcal{F}_\lambda &= \mathcal{A}_\lambda \int_1^\infty r^{1+2\mu} \phi(r;\lambda)f(r)\,dr.
\label{Flambda}
\end{align}
Here
\begin{align}
A_0 &= -2\ri\sqrt{\lambda_0}\,\frac{M_{\kappa_0,-\mu-1}(\zeta_0)}{W_{\kappa_0,\mu+1}(\zeta_0)}  \label{A0}
\\[6pt]
A_n &=-\frac{\ri}{\sqrt{\lambda_n}-\sqrt{\lambda_0}}\, \frac{\Gamma(1/2-\mu-\kappa_n)}{\Gamma(-2\mu)} \,
\frac{M_{\kappa_n,-\mu-1}(\zeta_n)}{\partial/\partial\lambda[W_{\kappa,\mu+1}(\zeta)]|_{\lambda=\lambda_n}}, \quad \mbox{($n=1,2,\ldots$)}   \label{An} \\[6pt]
\mathcal{A}_\lambda &=
\frac{\sqrt{\lambda}\,\exp[\nu^2\pi/2\sqrt{\lambda}]}{4\pi\mu^2(\lambda-\lambda_0)}\,
|W_{\kappa,\mu+1}(\zeta)|^{-2},  \label{Alambda}
\end{align}
with $\kappa_n$ and $\zeta_n$ being the values at $\lambda=\lambda_n$.

To put our dimensionless units in a solar context, we adopt the numbers quoted by \citet{bogcal95aa} for the $m=\frac{3}{2}$ polytrope with $\rho_0=1.96\times10^{-4}$ $\rm kg\,m^{-3}$, $p_0=6.20$ kPa, and $g=277.5$ $\rm km\,s^{-2}$: \emph{viz.}~$z_0=285$ km,  from which a frequency of 1 mHz corresponds to $\nu=0.247$ and 5 mHz to $\nu=1.23$. Dimensionless frequency $\nu=1$ corresponds to dimensional frequency 4.05 mHz. Frequencies $\nu\approx1$ are therefore seismically interesting, and sources placed at $s=2$ are only 285 km below the surface. For comparison, the detector resolution per pixel of the Helioseismic and Magnetic Imager (HMI) aboard the Solar Dynamics Observatory (SDO) is around 360 km.

%%%%%%%
\subsection{Special Cases: Orthogonality}
The expected orthogonality is easily confirmed. Specifically, with $f(s)=\psi_n(s)$ we find that $F_j=\delta_{jn}$ and $\mathcal{F}_\lambda=0$. Similarly, for $f(s)=\phi(s;\Lambda)$ it follows that $F_n=0$ for all $n$ and $\mathcal{F}_\lambda=\delta(\lambda-\Lambda)$ \citep[see][Appendix A]{HinJai12aa}.

%%%%%%%%
\subsection{Special Case: Delta Function}
\begin{figure}[htbp]
\begin{center}
\includegraphics[width=0.5\textwidth]{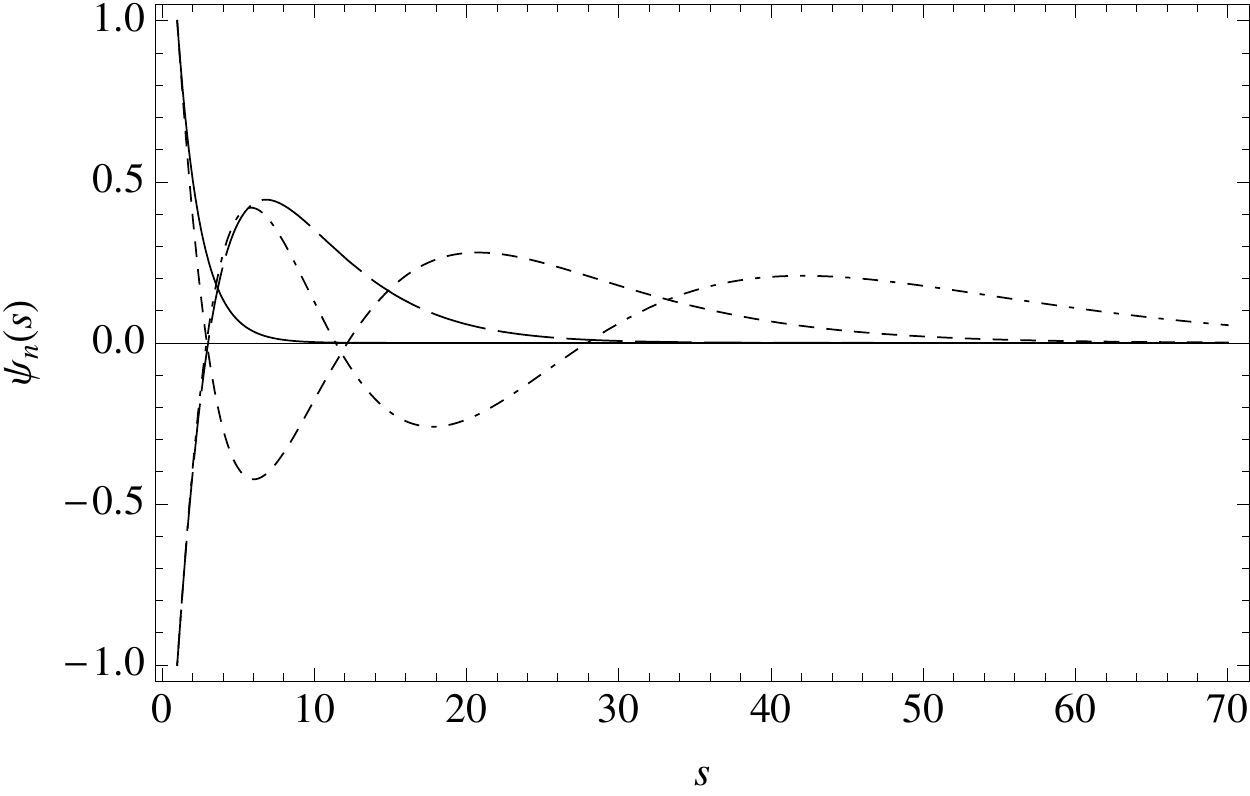}
\caption{The $f$- (full curve) and first three $p$-modes (respectively long dashed, short dashed, and dot-dashed) as functions of dimensionless depth for the case $\mu=\frac{1}{4}$, $\nu=1$. Higher order modes extend ever deeper rather than displaying finer structure. The modes have been normalized to $(-1)^n$ at the surface.}
\label{fig:pmodes}
\end{center}
\end{figure}

Unlike the familiar case of an oscillating string of finite length, arbitrarily fine spatial scales are not attained by the discrete spectrum of waves in a polytrope. This is because, as $n$ increases, the eigenfunctions push deeper rather than becoming more oscillatory (per unit length). This is illustrated in Figure \ref{fig:pmodes}. It is therefore impossible to construct an arbitrary structure from the $f$- and $p$-modes alone, particularly a structure of small extent.

The most extreme example is the Dirac delta function. With $f(s)=\delta(s-\sigma)$, ($\sigma>1$), Equation (\ref{expand}) yields
\begin{equation}
\delta(s-\sigma) =\sigma^{1+2\mu}\left( \sum_{n=0}^\infty A_n\psi_n(s)\psi_n(\sigma)+\int_0^\infty\mathcal{A}_\lambda\phi(s;\lambda)\phi(\sigma;\lambda)\,d\lambda\right). \label{delta}
\end{equation}
It is interesting to see how the continuous spectrum does this mathematically. Clearly the $\delta$-function requires indefinitely large wave numbers to synthesise its sharp peak. This implicates the large $\lambda$ behaviour of $\mathcal{A}_\lambda\phi(s;\lambda)\phi(\sigma;\lambda)$. It is easily shown that
\begin{multline}
\phi(s;\lambda) =
\left(2\mu s^{-1/2-\mu} +\mathcal{O}\left(\frac{1}{\sqrt{\lambda}}\right)\right)\cos\left[\sqrt{\lambda}(s-1)+\mathcal{O}\left(\frac{1}{\sqrt{\lambda}}\right) \right] %\\
+\mathcal{O}\left(\frac{1}{\sqrt{\lambda}}\right)\sin\left[\sqrt{\lambda}(s-1)+\mathcal{O}\left(\frac{1}{\sqrt{\lambda}}\right) \right] \quad
\label{phias}
\end{multline}
as $\lambda\to\infty$ \citep[using formulae from][Section 13.19]{DLMF}. In the same limit, $\mathcal{A}_\lambda\sim1/(4\pi\mu^2\sqrt{\lambda})$, so the dominant contribution to $\sigma^{1+2\mu}\mathcal{A}_\lambda\phi(s;\lambda)\phi(\sigma;\lambda)$ is
\begin{equation}
a_\lambda(s,\sigma)=\frac{1}{\pi\sqrt{\lambda}} \left(\frac{\sigma}{s}\right)^{1/2+\mu}\cos\sqrt{\lambda}(s-1)\cos\sqrt{\lambda}(\sigma-1).  \label{a}
\end{equation}
This is readily integrated using a simple change of variables and the standard result 
$
\int_0^\infty\cos kx \cos k \xi\, dk=\frac{\pi}{2}\delta(x-\xi)
$
to yield
\begin{equation}
\int_0^\infty a_\lambda(s,\sigma)\,d\lambda = \delta(s-\sigma).
\end{equation}
Hence
\begin{equation}
\int_0^\infty\left[ \mathcal{A}_\lambda\phi(s;\lambda)\phi(\sigma;\lambda)-\sigma^{-1-2\mu}a_\lambda(s,\sigma)\right]d\lambda + \sum_{n=0}^\infty A_n\psi_n(s)\psi_n(\sigma)=0.   \label{match_non_delta}
\end{equation}
This makes explicit how the continuous spectrum forms a $\delta$-function whilst also cancelling out the $f$- and $p$-modes.

Although the integral on the left hand side is computationally problematic, Equation (\ref{match_non_delta}) is confirmed  numerically to several significant figures. To avoid extreme round-off error and consequent catastrophic loss of precision it is best to integrate the two terms in square brackets separately. The integrals may be truncated, $\lambda<\lambda_{\rm max}$, at sufficiently large $\lambda_{\rm max}$ since the two terms are barely distinguishable there. $\lambda_{\rm max}=5000$ is generally adequate. The discrete sum can also be very slowly converging in practice, but is easily accelerated using repeated Shanks transformation \citep{BenOrs78aa}. In practice, using $n=0,\ldots,20$ with ten Shanks transforms works as well as retaining 120 or more terms without transformation, and is much quicker to evaluate.

%%%%%%%%
\subsection{Special Case: Gaussian}
\begin{figure}[tbhp]
\begin{center}
\includegraphics[width=0.5\textwidth]{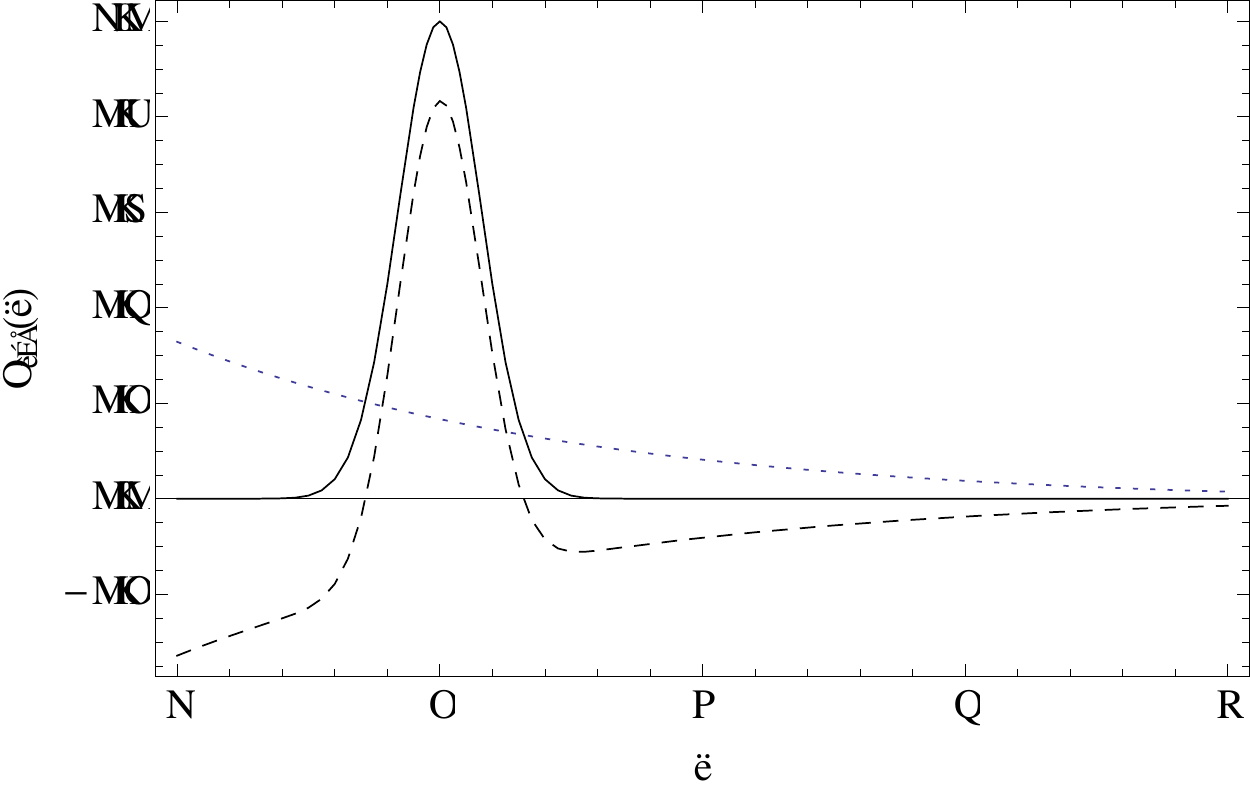}
\caption{The reconstructed Gaussian $g(s)=\exp[-20(s-2)^2]$ (full curve, indistinguishable from the original function) together with the discrete (dotted, dominated by the $f$-mode) and continuum (dashed) contributions (see Equation (\ref{expand})).}
\label{fig:gaussian}
\end{center}
\end{figure}

We next consider a Gaussian $g(s)=\exp[-20(s-2)^2]$. This is numerically simpler than the $\delta$-function case, as the integral in Equation (\ref{expand}) converges in the normal sense. Figure \ref{fig:gaussian} displays the reconstructed Gaussian, as well as the discrete and continuum contributions. Once again, as for the $\delta$-function case, the sharp peak is contributed by the continuous spectrum. There is no hint of it in the discrete sum of $f$- and $p$-modes.

%%%%%%%%%%%%%%%%%%%%%%%%%%%%%%%
\break\section{Constructing a Wave Field with a Forcing Term} 
\subsection{Forced Wave Equation and Solutions}
The above mathematical development may be used to construct the full wave field $\Phi(\x_\perp,s)$ in multiple dimensions. Let $x$ denote the horizontal coordinate in a 2D cartesian setting, and $r$ the usual radial coordinate in an axially symmetric scenario. Both are assumed to have already been scaled by $z_0$, and are therefore dimensionless. Consider a forcing term at $x=0$ or $r=0$ on the right hand side of the wave equation (assuming an $e^{-i\nu\tau}$ time dependence as usual, where $\tau=t\sqrt{g/m z_0}$ is dimensionless time),
\begin{equation}
\left(\tilde\nabla^2+\frac{m}{s}\pderiv{}{s}+\frac{\nu^2}{s}\right)\Phi(\x_\perp,s) =
f(s)\delta(\x_\perp)=
\begin{cases}
f(s)\delta(x) & \text{(2D cartesian)}\\[10pt]
\pi^{-1}r^{-1}f(s)\delta(r) & \text{(cylindrical)}
\end{cases}.
\label{waveeqn3Ddriv}
\end{equation}
Only axisymmetric drivers are considered in the cylindrical case. The forced solutions may be written in terms of the discrete and continuous spectra:
\begin{equation}
\Phi_{\rm f}(x,s,\tau) = -\half\,e^{-\ri\nu\tau}\left[\sum_{n=0}^\infty \frac{i}{\sqrt{-\lambda_n}}\,e^{\ri\sqrt{-\lambda_n}\,|x|} F_n\, \psi_n(s) + \int_0^\infty \frac{1}{\sqrt{\lambda}}\,e^{-\sqrt{\lambda}\,|x|}\mathcal{F}_\lambda \, \phi(s;\lambda)\,d\lambda \right]  \label{expand2Dfcart}
\end{equation} 
or
\begin{equation}
\Phi_{\rm f}(r,s,\tau) = -e^{-\ri\nu\tau}\left[\frac{i}{4}\sum_{n=0}^\infty H_0^{(1)}\left(\sqrt{-\lambda_n} \,r\right) F_n\, \psi_n(s) + \frac{1}{2\pi}\int_0^\infty K_0\left(\sqrt{\lambda}\,r\right)\mathcal{F}_\lambda \, \phi(s;\lambda)\,d\lambda \right]  ,\label{expand2Dfcyl}
\end{equation}
where $H_0^{(1)}=J_0+i\,Y_0$ is the Hankel function of the first kind of order 0 (representing an outward travelling wave) and $K_0$ is the modified Bessel function of the second kind (representing the jacket). Both are logarithmically singular on $r=0$. (The factors multiplying the sum and integral in Equation (\ref{expand2Dfcyl}) may be derived by integrating Equation (\ref{waveeqn3Ddriv}) over a disk of arbitrary radius and applying the divergence theorem and several Bessel function identities.) The coefficients $F_n$ and $\mathcal{F}_\lambda$ are as given by Equations (\ref{Fn}--\ref{Flambda}). Written in these forms, the continuum contribution is real and the discrete contribution is complex. For the axisymmetric case there is a logarithmic singularity at $r=0$ in both the continuum term and the real part of the discrete term. The imaginary part of the discrete term is not singular.

An alternate construction, where $\Phi$ or $\partial\Phi/\partial s$ is prescribed directly on the axis, is discussed briefly in Appendix \ref{SausKink}.

\begin{figure}[tbbp]
\begin{center}
\centerline{
\hfill
\includegraphics[width=0.48\textwidth]{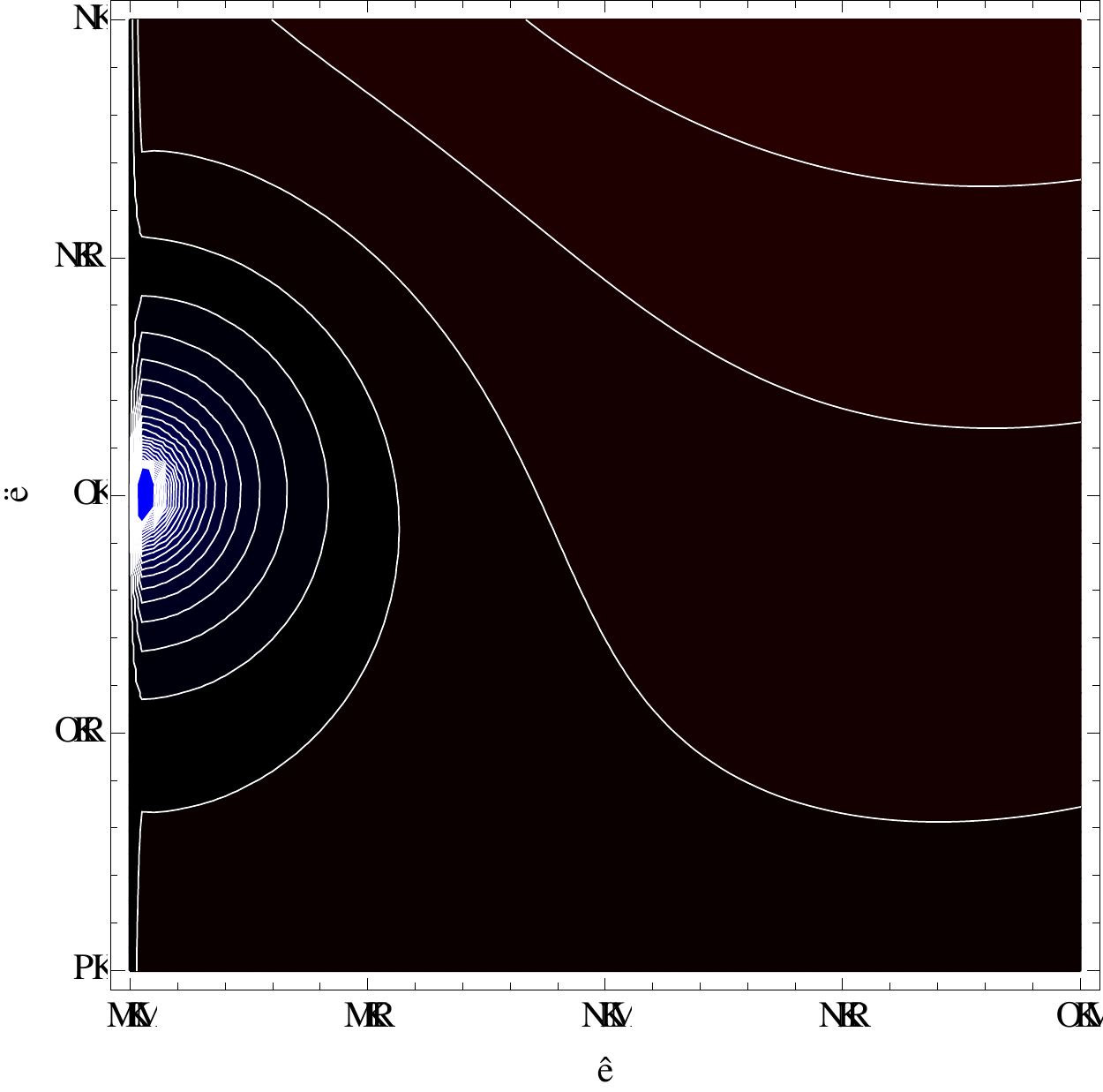}\hfill
\includegraphics[width=0.48\textwidth]{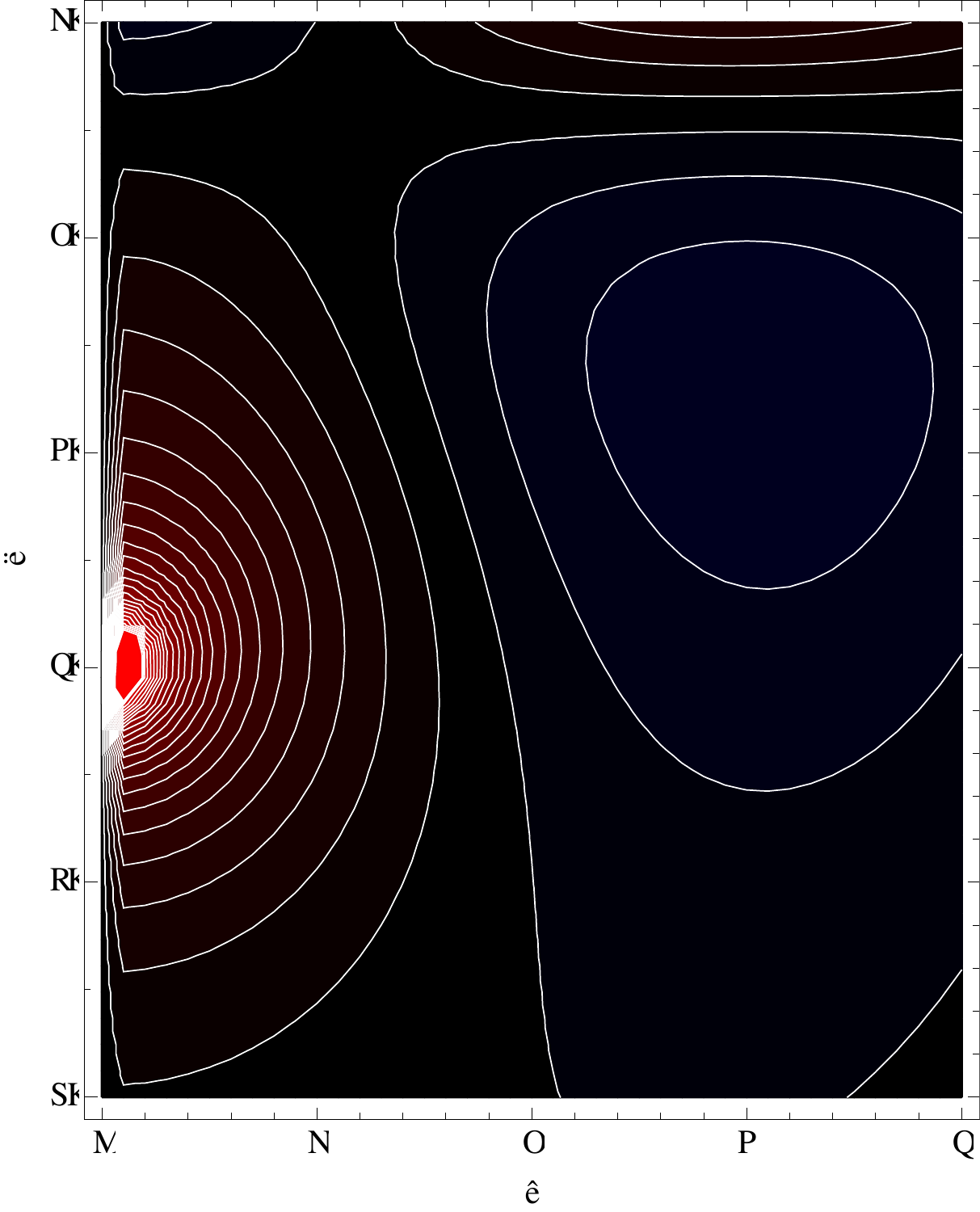}\hfill}
\caption{Animations (contour plots of $\Phi_{\rm f}$) for the axisymmetric point source with $\nu=1$, $\sigma=2$ (left) and $\nu=2$, $\sigma=4$ (right), illustrating the generation of respectively the $f$-mode by a shallow source and the $p_1$ mode by a deeper source. Blue and red shading indicates different signs of $\Phi$. The colour maps and contour sequences are clipped near the source. Run with `Loop' set. Files: Deltad\_1\_2.mp4 and Deltad\_2\_4.mp4.}
\label{fig:movie}
\end{center}
\end{figure}

We now restrict attention to the axisymmetric point source, $f(s)=\delta(s-\sigma)$, $\sigma>1$. The $\delta$-function case is rather tricky numerically. It is again best to extract the large-$\lambda$ behaviour from the integrand and sum it exactly in the sense of generalized functions. To that end we rewrite the continuum integral (with $\mathcal{F}_\lambda=\sigma^{1+2\mu}\mathcal{A}_\lambda \phi(\sigma;\lambda)$) in the form
\begin{multline}
\int_0^\infty K_0(\sqrt{\lambda}\,r)\mathcal{F}_\lambda \, \phi(s;\lambda)\,d\lambda = 
\frac{1}{2} \left(\frac{1}{\sqrt{r^2+(s+\sigma
   -2)^2}}+\frac{1}{\sqrt{r^2+(s-\sigma )^2}}\right)
   \left(\frac{\sigma }{s}\right)^{\mu +\frac{1}{2}}
 \\[6pt]
   +\lim_{\lambda_{\rm max}\to\infty}\left[\sigma^{1+2\mu} \int_0^{\lambda_{\rm max}} K_0(\sqrt{\lambda}\,r)\mathcal{A}_\lambda \, \phi(\sigma;\lambda)\, \phi(s;\lambda)\,d\lambda 
 -\int_0^{\lambda_{\rm max}} K_0(\sqrt{\lambda}\,r) a_\lambda(s,\sigma)\,d\lambda\right] 
 , \quad r\ne0,\quad         
 \end{multline}
where the first term on the right hand side is the exact limit of the second integral in square brackets. It's form indicates that the dominant high-wavenumber contribution to the continuum integral is simply radial inverse-distance falloff from the source at $(0,\sigma)$ and its image at $(0,2-\sigma)$. As before, $a_\lambda$ is given by Equation ({\ref{a}). Then both the first and second integrals need only be integrated numerically over $0<\lambda<\lambda_{\rm max}$ for sufficiently large but in practice moderate $\lambda_{\rm max}$ ($(10/r)^2$ is typically more than adequate), since their integrands nearly cancel at large $\lambda$. This device significantly reduces numerical noise at small $r$, but is unnecessary at large $r$ where the decaying Bessel-$K_0$ accelerates convergence.

Figure \ref{fig:movie} and accompanying animations illustrate the generation of $f$- and $p_1$ modes by axisymmetric $\delta$-function sources of two depths (and frequencies; the second panel corresponds to a higher frequency so as to make the $p_1$ mode shallower for graphical purposes).

%%%%
\subsection{Surface Velocity}\label{surf}
From a seismic perspective, we may be most interested in the observed vertical component of velocity at the surface, $v_z=-i\,\nu\, \partial\Phi/\partial s$ (where a dimensional scaling $\sqrt{g z_0/m}$ has been removed). Since $\partial\Phi/\partial s=-(\nu^2/m)\Phi$ at the surface by boundary condition (\ref{BC1}), we may identify $v_z$ with $\Phi$ there, apart from a normalization.

The contour plots in Figure \ref{fig:movie} and accompanying animations suggest that the $f$-mode is preferentially excited by shallow sources, and the $p$-modes by deeper excitation, presumably because the $f$-mode eigenfunction is dominant near the surface, with the $p$-modes residing considerably deeper (see Figure \ref{fig:pmodes}). This poses the question of which modes precisely are excited by a frequency $\nu$ point source at depth $\sigma$. At sufficiently large radius (beyond the jacket) the surface velocity power is dominated by the discrete spectrum and the Hankel functions become essentially complex exponentials (with an $r^{-1/2}$ amplitude factor), so we have the usual Fourier orthogonality in the horizontal plane, allowing us to write velocity power as\footnote{The $| H_0^{(1)}(k_n\,r)|^2$ term decays with radius as $2\,\pi^{-1}k_n^{-1}r^{-1}$ for sufficiently large $r$ .}
\begin{equation}
P(r) \propto \frac{1}{r}\sum_{n=0}^\infty \frac{1}{\sqrt{-\lambda_n}} \, \left| F_n\psi_n(1)\right|^2\,.
\end{equation}
This may be summed, and the fractions in each of the modes $n$, with $n=0$ corresponding to the $f$-mode and $n>0$ to $p_n$, may be calculated for various frequencies and source depths. The results are summarised in Figure \ref{fig:powergridd}. %Raw power is plotted in Figure \ref{fig:pt1Raw} for $\nu=1$, showing a strong weighting towards the $f$-mode for shallow sources.

\begin{figure}[tbhp]
\begin{center}
\includegraphics[width=0.9\textwidth]{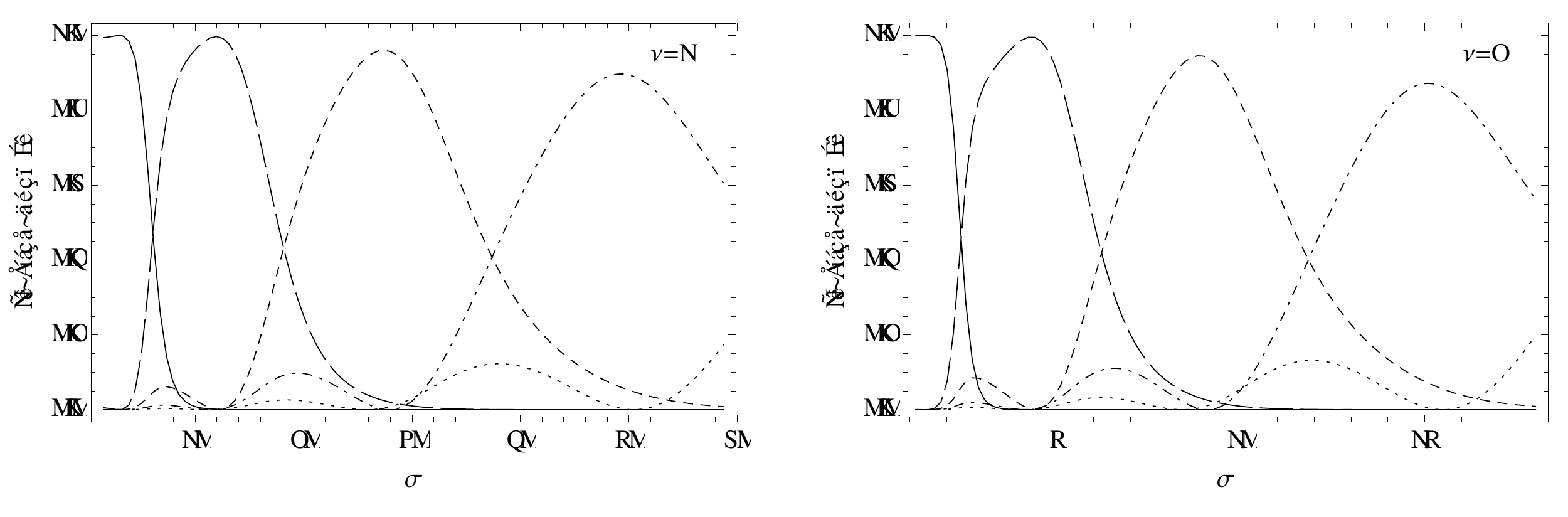}
\caption{Fractional surface power in the $f$ (full curve), $p_1$ (long dash), $p_2$ (short dash), $p_3$ (dot-dashed), and $p_4$ (dotted) modes excited by a harmonic axisymmetric $\delta$-function source of frequency $\nu$ placed at depth $\sigma$. Left: $\nu=1$; right: $\nu=2$. In each graph, the $\sigma$ is arbitrarily chosen to range down to the $p_4$ Lamb depth, $s_L=-\nu^2/\lambda_4$. We see that fractional power is almost exclusively dependent on $n$ and $\sigma/s_L$ alone.}
\label{fig:powergridd}
\end{center}
\end{figure}

%\begin{figure}[tbhp]
%\begin{center}
%\includegraphics[width=0.5\textwidth]{pt1Raw}
%\caption{Total surface power in the $f$ (full curve), $p_1$ (long dash), $p_2$ (short dash), $p_3$ (dot-dashed), and $p_4$ (dotted) modes excited by a harmonic axisymmetric $\delta$-function source of frequency $\nu$ placed at depth $\sigma$ for $\nu=1$}
%\label{fig:pt1Raw}
%\end{center}
%\end{figure}

The dependence on frequency may be understood by recourse to the complete polytrope, which extends all the way up to $z=0$ and is a pure power law with no length scale. Its discrete eigenvalues are exactly $\omega^2=gk(1+2n/m)$. The depth dependence of its eigenfunctions therefore scales exactly with $k^{-1}$ and hence with $\omega^{-2}$. Consequently, $p$-mode excitation as a function of point source depth can depend only on depth through $kz$ or equivalently $\omega^2z$. For the truncated polytrope this is only approximately true. Nevertheless, Figure \ref{fig:powergridd} indicates that it is a good approximation.

Figure \ref{fig:VzGrid} depicts surface $v_z$ as a function of radius for four source depths of frequency $\nu=1$. It is clear that the acoustic jacket at the surface extends ever farther as the source moves deeper. In each case there is of course a very compact logarithmic singularity in both the discrete and continuum velocities, but these cancel exactly in the full solution, since the current case is equivalent to specifying $2\pi\Phi(0,s,\tau)/\ln r\to f(s)\,e^{-i\nu\tau}$ as $r\to0$ (see Appendix \ref{SausKink}), which vanishes at the surface for $f(s)=\delta(s-\sigma)$. This does not mean that $\Phi$ vanishes at $r=0$, $s=1$, only that the singular part does. That leaves only nonsingular components. We refer to this as `healing' the logarithmic singularity, though the $\sigma=1.2$ case in Figure \ref{fig:VzGrid} illustrates that the healing can still leave a sharp velocity peak at $r=0$ if the source is too shallow, as one might expect. The removal of the logarithmic singularity still leaves a broader continuum jacket as is apparent from the grey curve in the figure. Apart from the very shallow source case, it is interesting that even this `outer' jacket seems to largely counteract a similar feature in the real part of the discrete solution, further hiding the source from view. In the 2D cartesian case (Equation (\ref{expand2Dfcart})), where there is no singularity, only this regular part of the jacket exists.

\begin{figure}[tbhp]
\begin{center}
\includegraphics[width=.89\textwidth]{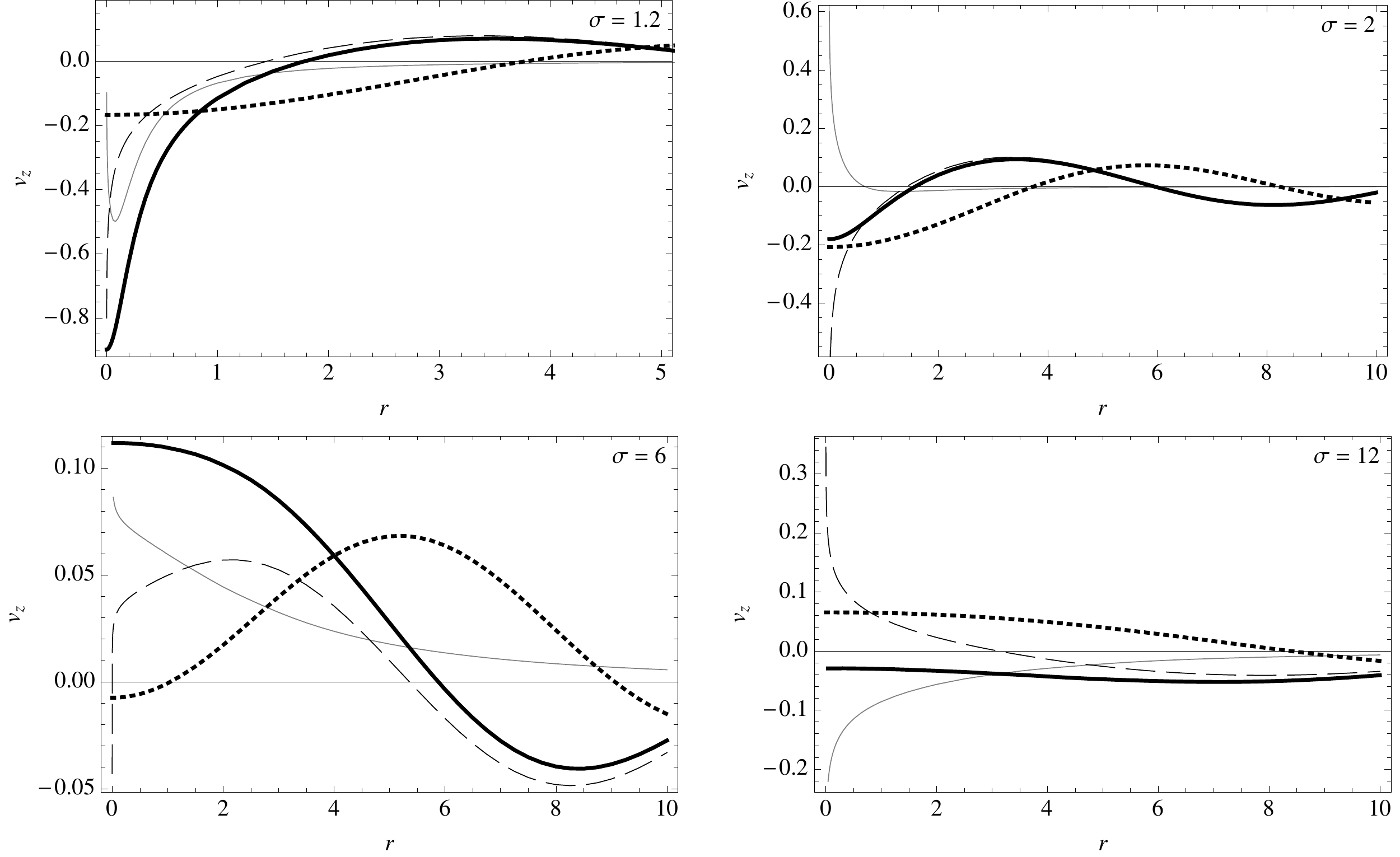}
\caption{Surface vertical velocity $v_z$ against radius $r$ for $\delta$-function sources at four different depths: $\sigma=1.2$ (top left); $\sigma=2$ (top right); $\sigma=6$ (bottom left); and $\sigma=12$ (bottom right). Note that the $\sigma=1.2$ case is depicted over a reduced radius to better display its behaviour near the origin. In each case the thick black full and dotted curves represent real and imaginary parts of the full solution. The grey curve represents the continuum velocity (which is purely real). The dashed curve is the real part of the discrete component (the imaginary part is again the thick dotted curve).}
\label{fig:VzGrid}
\end{center}
\end{figure}

\begin{figure}[tbhp]
\begin{center}
\includegraphics[width=0.89\textwidth]{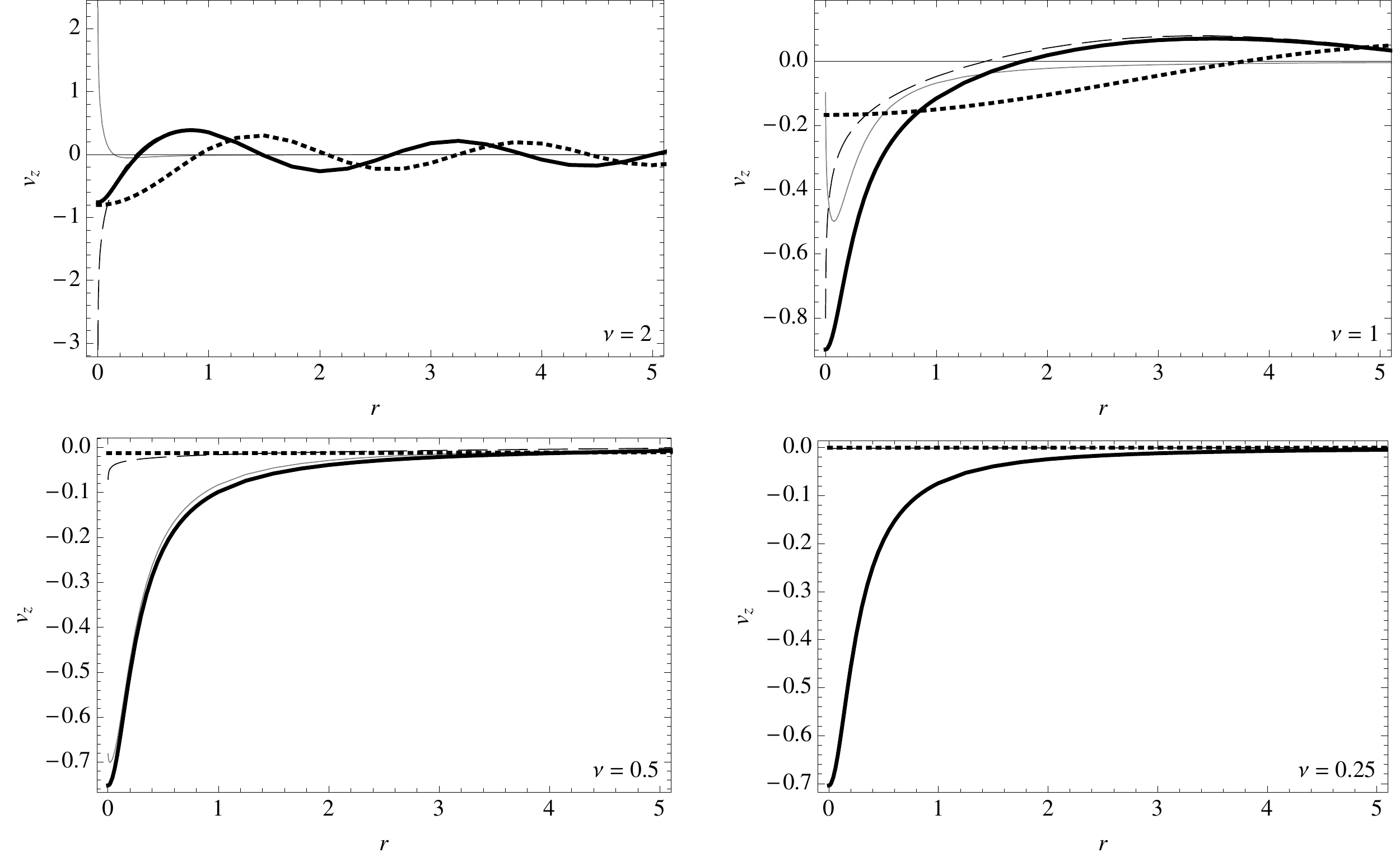}
\caption{As for the $\sigma=1.2$ case of Figure \ref{fig:VzGrid}, but with frequencies $\nu=2$ (top left), $\nu=1$ (top right), $\nu=0.5$ (bottom left) and $\nu=0.25$ (bottom right). In each case the thick black full and dotted curves represent real and imaginary parts of the full solution respectively. The grey curve represents the continuum velocity. The dashed curve is the real part of the discrete component.}
\label{fig:lofreq}
\end{center}
\end{figure}

The dependence of surface signature on frequency is illustrated in Figure \ref{fig:lofreq} for a shallow point source at $\sigma=1.2$, \emph{i.e.}~only some 57 km below the surface. At high frequency $\nu=2$ (8.1 mHz) the discrete spectrum dominates since the jacket is extremely narrow (grey curve), and merely cancels the singularity in the discrete solution (dashed). On the other hand, small-$r$ surface velocities are progressively dominated by the continuous spectrum as $\nu$ diminishes. Since the dominant behaviour $a_\lambda$ of that contribution is independent of $\nu$ (see Equation (\ref{a})), we expect only weak dependence on frequency of $v_z$ at small $r$ as $\nu\to0$. This is illustrated by the $\nu=0.5$ (2 mHz) and 0.25 (1 mHz) cases in Figure \ref{fig:lofreq} being very similar. In both instances, there is a prominent and compact surface velocity feature immediately above the source. It is produced by the continuum (grey curve), suggesting that shallow low frequency jackets are potentially visible. The near-field is made more apparent at low frequencies because of the very low amplitude of the far-field. As the source is moved deeper, the ability of the continuum to heal and hide comes to the fore, and the surface `bump' becomes much less prominent.

%The $| H_0^{(1)}(k_n\,r)|^2$ term decays with radius as $2\,\pi^{-1}k_n^{-1}r^{-1}$ for sufficiently large $r$ (utilising the asymptotic behaviour of the Hankel function), representing a simple geometric attenuation. The $|K_0(\sqrt{\lambda}\,r)|^2$ term similarly decays as $\half\lambda^{-1/2}r^{-1} e^{-2\sqrt{\lambda}\,r}$, dominated by the jacket.

Despite the acoustic jacket being composed of a continuum of horizontal wavenumbers, its effect at the surface is to heal the logarithmic singularity and make further relatively minor changes to $v_z$ within a small radius about the source axis. Nevertheless, there may still be some measure of continuum power observable that shows up in the seismic power maps. To look for this we run a `slit' across the disk through the axis of the source, and compute the vertical velocity power spectrum for two source depths (Figure \ref{fig:WPower}). Using HMI-like per pixel spatial resolution $\Delta r=1.27$, there is no sign of any significant inter-ridge power in either case. The sources have very effectively hidden themselves.

\begin{figure}[tbhp]
\begin{center}
\includegraphics[width=0.9\textwidth]{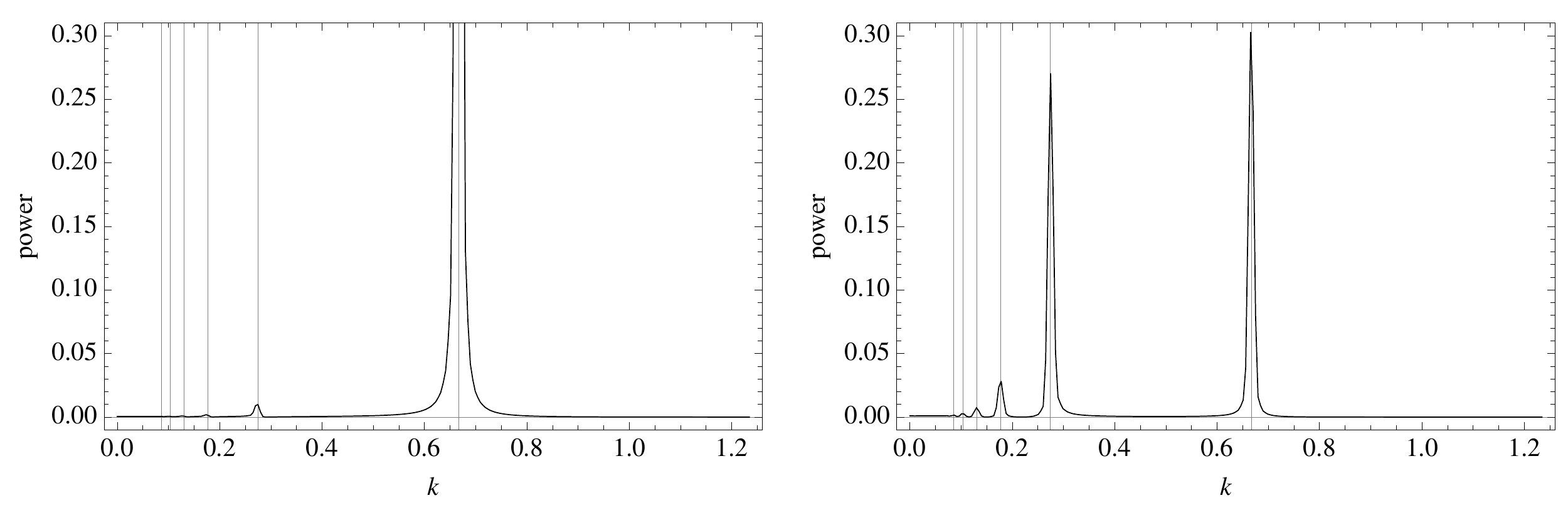}
\caption{Surface velocity power spectra of ${\sqrt[4]{r^2+1}}\,v_z$ along a slit through the source axes for $\nu=1$ sources at $\sigma=2$ (left) and $\sigma=6$ (right). The factor ${\sqrt[4]{r^2+1}}$ removes the $1/\sqrt{r}$ geometric amplitude attenuation without killing off the signal at $r=0$. We use a spatial grid of 1024 points of spacing $\Delta r=1.27$ (comparable to HMI resolution) and apply Hann windowing to suppress spurious side lobes. The vertical grey lines represent the eigen-wavenumbers of the $f$- and first five $p$-modes (right to left). As we saw earlier, the shallow source overwhelmingly excites the $f$-mode (peak power around 2.4 at this resolution), whereas the deeper source excites substantial $f$ and $p_1$ power, and even noticeable $p_2,\ldots p_4$. No significant background power is apparent. The $f$-mode peak is cut off for graphical purposes in the left panel, but extends up to about 2.4 at this resolution. The normalization of the two sources is the same.}
\label{fig:WPower}
\end{center}
\end{figure}

%%%%%%%%%%%%%%%%%%%%%%%%%%%%%%%
\section{Conclusions}
The acoustic continuum is an intrinsic component of the Sun's interior wave field, particularly for small scale sources not resolvable by summation over the discrete modes. We have summarized the theory of \citet{bogcal95aa} and applied it to constructing compact sources, in particular point sources. This has allowed us to visualise the role of the near-field in generating $f$- and $p$-modes, and to quantify the relative proportions of surface velocity power going into each of the discrete modes from monochromatic point sources of varying frequency and depth. As expected, shallow sources preferentially excite the $f$-mode, with subsequent $p$-modes taking precedence as the source is moved deeper or the frequency increased. 

The significance of the continuum `acoustic jacket' for observationally accessible surface velocities is explored in Section \ref{surf}. Within a small radius determined by source depth, the continuum plays a significant role. It heals the coordinate singularity at $r=0$, but does so with (of course) a continuum of wavenumbers. (In the real Sun, compact acoustic sources will have some non-zero spatial extent, possibly hundreds of km or more, making the logarithmic singularity irrelevant, but the `outer' jacket will persist.) The resulting surface $v_z$ displays a simple sinusoidal behaviour that is basically unremarkable, except if the source is very shallow indeed (see the $\sigma=1.2$ case in Figure \ref{fig:VzGrid} and the low frequency cases of Figure \ref{fig:lofreq}). The effects of the continuous spectrum for buried point sources do not contribute substantially to background power in the surface doppler power spectrum for the cases considered.

The fact that the wave field close to the source beneath the surface is dominated by the jacket modes must surely place limits on how well reconstructive techniques such as Acoustic Holography \citep{LinBra00aa} can image them. We have not attempted this, but the optical assumptions underlying the method clearly break down within the jacket, irrespective of any further finite wavelength effects.

%In summary, we have shown how the continuous spectrum of compact acoustic sources in the Sun is quite adept at hiding the source.

%%%%%%%%%%%%%%%%%%%%%%
%%%%%%%%%%%%%%%%%%%%%%%%%%%%%%%
\appendix
\section{Wave Field Matching to a Prescribed Sausage or Kink Oscillation}
\label{SausKink}
An alternate scenario relevant to a thin vertical magnetic flux tube at $r=0$ is to directly prescribe the pressure perturbation $\Phi$ (the axisymmetric sausage mode) or the radial displacement $\partial\Phi/\partial r$ (the kink wave, with an $e^{\pm i\theta}$ azimuthal dependence) on the axis. For the kink, with $2\pi r^2\partial\Phi/\partial r\to f(s)\,e^{i(\pm\theta-\nu\tau)}$ as $r\to0$,
\begin{multline}
\Phi_{\rm k}(r,\theta,s,\tau) = -e^{\ri(\pm\theta-\nu\tau)}\left[\frac{i}{4}\sum_{n=0}^\infty \sqrt{-\lambda_n}\,H_1^{(1)}\left(\sqrt{-\lambda_n} \,r\right) F_n\, \psi_n(s) +\right. \\
\left. \frac{1}{2\pi}\int_0^\infty \sqrt{\lambda}\,K_1\left(\sqrt{\lambda}\,r\right)\mathcal{F}_\lambda \, \phi(s;\lambda)\,d\lambda \right]  ,\label{expandkink}
\end{multline}
where we have used $H_{-1}^{(1)}(z)=-H_1^{(1)}(z)$ and $K_{-1}(z)=K_1(z)$.
For the sausage mode with $2\pi\Phi(0,s,\tau)/\ln r\to f(s)\,e^{-i\nu\tau}$ as $r\to0$,
\begin{equation}
\begin{split}
\Phi_{\rm s}(r,s,\tau) &= -e^{-\ri\nu\tau}\left[\frac{i}{4}\sum_{n=0}^\infty H_0^{(1)}\left(\sqrt{-\lambda_n} \,r\right) F_n\, \psi_n(s) + \frac{1}{2\pi}\int_0^\infty K_0\left(\sqrt{\lambda}\,r\right)\mathcal{F}_\lambda \, \phi(s;\lambda)\,d\lambda \right]  \\[6pt]
&= \Phi_{\rm f}(r,s,\tau)\,.
\end{split}   \label{expandsaus}
\end{equation}
We see that the axisymmetric forced case and the sausage case are equivalent. We will not pursue the kink case here, except to note that the added factor of $\sqrt{\lambda}$ in the integrand emphasises higher wavenumbers, meaning that the kink jacket is more tightly bound to the source.

%%%%%%%
\acknowledgments
The author thanks Sergey Zharkov for asking the question.

%%%%%%%%%%%%%%%%%%%%%%%%%%%%%%%
%  REFERENCES

\bibliographystyle{apj}        
%\bibliography{fred}
\bibliography{jackets_rev}

\end{document}